\magnification \magstep1
\raggedbottom
\openup 2\jot
\voffset6truemm
\pageno=1

\def\cstok#1{\leavevmode\thinspace\hbox{\vrule\vtop{\vbox{\hrule\kern1pt
\hbox{\vphantom{\tt/}\thinspace{\tt#1}\thinspace}}
\kern1pt\hrule}\vrule}\thinspace}

\leftline {\bf PUSH ON A CASIMIR APPARATUS IN A WEAK GRAVITATIONAL FIELD}
\vskip 1cm
\leftline{\it Giuseppe Bimonte, Enrico Calloni, Giampiero Esposito, Luigi Rosa}
\vskip 1cm
\noindent
Dipartimento di Scienze Fisiche, Complesso Universitario
di Monte S. Angelo, Via Cintia, Edificio N', 80126 Napoli, Italy. 
\vskip 0.3cm
\noindent
INFN, Sezione di Napoli, Complesso Universitario di 
Monte S. Angelo, Via Cintia, Edificio N', 80126 Napoli, Italy.
\vskip 1cm
\noindent
{\bf Abstract}. The influence of the gravity acceleration 
on the regularized energy-momentum
tensor of the quantized electromagnetic field between
two plane parallel conducting plates is
derived. We use Fermi coordinates and work to first order in the
constant acceleration parameter. A new simple 
formula for the trace anomaly is found to first order in the constant
acceleration, and a more systematic derivation is therefore obtained of
the theoretical prediction according to which the Casimir device in a weak
gravitational field will experience a tiny push in the upwards direction.
\vskip 100cm
\leftline {\bf 1. Introduction}
\vskip 0.3cm
\noindent
An important property of quantum electrodynamics is
that suitable differences of zero-point energies of the quantized
electromagnetic field can be made finite and produce measurable effects
such as the tiny attractive force among perfectly conducting parallel
plates known as the Casimir effect [1]. This is a remarkable
quantum mechanical effect that makes itself manifest on a macroscopic
scale. For perfect reflectors and metals the Casimir force can be
attractive or repulsive, depending on the geometry of the cavity, whereas
for dielectrics in the weak-reflector approximation it is always attractive,
independently of the geometry [2]. The Casimir effect can be
studied within the framework of boundary effects in quantum field theory,
combined with zeta-function regularization or Green-function methods,
or in more physical terms, i.e. on considering van der Waals forces
[3] or scattering problems [4]. Casimir energies
are also relevant in the attempt of building a quantum theory of gravity
and of the universe [5].

For these reasons, in Ref. [6] we evaluated the force
produced by a weak gravitational field on a rigid Casimir cavity.
Interestingly, the resulting force was found to have opposite
direction with respect to the gravitational acceleration;
moreover, we found that the current experimental sensitivity of
small force macroscopic detectors would make it possible, at least
in principle, to measure such an effect [6]. More
precisely, the gravitational force on the Casimir cavity might be
measured provided one were able to use rigid cavities and  find an
efficient force modulation method [6]. Rigid cavities,
composed by metal layers separated by a dielectric layer, make it
possible to reach separations as small as 5$\div$10 nm and allow
to build multi-cavity structures, made by a sequence of such
alternate layers. If an efficient modulation method could be
found, it would be possible  to achieve a modulated force of order
$10^{-14}$ N in the earth's gravitational field. The measure of
such a force, already possible with current small-force detectors
on macroscopic bodies, might open the way to the first test of the
gravitational field influence on vacuum energy [6]. In
Ref. [6], calculations were based on simple assumptions
and the result can be viewed as a reasonable ``{\it first order}''
generalization of $T_{\mu\nu}$ from Minkowski to curved
space-time. The present paper is devoted to a deeper understanding
and to more systematic calculations of the interaction of a weak
gravitational field with a Casimir cavity. To first order in our
approximation the former value of the force exerted by the field
on the cavity is recovered. But here we also find a trace anomaly
for the energy-momentum tensor.

We consider a plane-parallel Casimir cavity, made of
ideal metallic plates, at rest in the gravitational field
of the earth, with its plates lying in a horizontal plane.
We evaluate the influence of the gravity acceleration
$g$ on the Casimir cavity but neglect any variation
of the gravity acceleration across the cavity, and therefore
we do not consider the influence of tidal forces.
The separation $a$ between the plates is taken to be much
smaller than the extension of the plates, so that edge
effects can be neglected. We obtain a perturbative
expansion of the energy-momentum tensor of the electromagnetic field
inside the cavity, in terms of the small parameter
$\epsilon \equiv 2 ga/c^2$, to first order in $\epsilon$. For this purpose,
we use a Fermi [7,8] coordinates system
$(t,x,y,z)$ rigidly connected to the cavity.
The construction of these coordinates
involves only invariant quantities such as the observer's proper time,
geodesic distances from the world-line, and components of tensors
with respect to a tetrad [8]. This feature makes it possible
to obtain a clear
identification of the various terms occurring in the metric.
In our analysis we adopt the covariant point-splitting procedure
[9,10] to compute the perturbative expansion
of the relevant Green functions. Gauge invariance
plays a crucial role and we check it up to first order by means of
the Ward identity. We also evaluate the Casimir energy and pressure, and
in this way we obtain a sound derivation of the result in Ref. [6],
according to which the Casimir device in a weak gravitational field
will experience a tiny push in the upwards direction. 

Use is here made of mixed
boundary conditions on the potential plus Dirichlet conditions
on ghost fields. With our notation,
the $z$-axis coincides with the vertical upwards direction, while the
$(x,y)$ coordinates span the plates, whose equations are $z=0$ and
$z=a$, respectively. The resulting line element for a non-rotating system
is therefore [7]
$$
ds^{2}= -c^{2} \left(1+\epsilon {z\over a} \right)dt^{2}
+dx^{2}+dy^{2}+dz^{2} + {\rm O}(|x|^{2}) 
= \eta_{\mu \nu}dx^{\mu}dx^{\nu}-\epsilon {z\over a}
c^{2} dt^{2},
\eqno (1.1)
$$
where $\eta_{\mu \nu}$ is the flat Minkowski metric
${\rm diag}(-1,1,1,1)$.
\vskip 10cm
\leftline {\bf 2. Green Functions}
\vskip 0.3cm
\noindent
For any field theory, once that the invertible differential operator
$U_{ij}$ in the functional integral is given, the corresponding Green
functions satisfy the condition (we use hereafter the DeWitt 
condensed-index notation)
$$
U_{ij}G^{jk}=-\delta_{i}^{\; k},
\eqno (2.1)
$$
and are boundary values of holomorphic functions. The choice of boundary
conditions will determine whether we deal with advanced Green functions
$G^{+jk}$, for which the integration contour passes below the poles of the
integrand on the real axis, or retarded Green functions $G^{-jk}$, for
which the contour passes instead above all poles on the real axis, or yet
other types of Green functions. Among these, a key role is played by the
Feynman Green function $G_{F}^{jk}$, obtained by choosing a contour that
passes below the poles of the integrand that lie on the negative real axis
and above the poles on the positive real axis. If one further defines
the Green function [11]
$$
{\overline G}^{jk} \equiv {1\over 2}(G^{+jk}+G^{-jk}),
\eqno (2.2)
$$
one finds in stationary backgrounds (for which the metric is independent of
the time coordinate, so that there exists a timelike Killing vector field)
that the Feynman Green function has a real part equal to
${\overline G}^{jk}$, and an imaginary part equal to the Hadamard function
$H^{jk}$, i.e.
$$
H^{jk}(x,x') \equiv -2i \Bigr[G_{F}^{jk}(x,x')
-{\overline G}^{jk}(x,x')\Bigr].
\eqno (2.3)
$$
This relation can be retained as a definition when the background is
nonstationary; in such a case, however, $H^{jk}(x,x')$ is generally
no longer real [11].

In particular, the photon Green function $G_{\lambda \nu'}$ in a curved
spacetime with metric $g_{\mu \nu}$ solves the equation [12]
$$
\sqrt{-g}P^{\; \lambda}_\mu(x) G_{\lambda\nu'}=g_{\mu\nu}
\delta(x,x').
\eqno (2.4)
$$
The wave operator $P_{\mu}^{\; \lambda}$ results from the gauge-fixed
action with Lorenz [13] gauge-fixing functional
$\Phi_{L}(A) \equiv \nabla^{\mu}A_{\mu}$, and having set to $1$ the
gauge parameter of the general theory, so that 
$$
P^{\; \lambda}_\mu(x)=-\delta^{\; \lambda}_\mu \cstok{ }_x
+R^{\; \lambda}_\mu(x),
\eqno (2.5)
$$
where $\cstok{ }_x \equiv g^{\alpha\beta}
\nabla_{\alpha}\nabla_{\beta}(x)$.
Since we need the action of the gauge-field operator
$P^{\; \lambda}_\mu(x)$ on the photon Green functions,
it is worth noticing that
$$
D_{\beta\lambda\nu'}  \equiv  \nabla_\beta
G_{\lambda\nu'} = \partial_\beta
G_{\lambda\nu'}-\Gamma^\mu_{\beta\lambda}G_{\mu\nu'},
\eqno (2.6)
$$
$$
Q_{\alpha\beta\lambda\nu'}  \equiv \nabla_\alpha \nabla_\beta
G_{\lambda\nu'} = \nabla_\alpha D_{\beta\lambda\nu'}
=\partial_\alpha D_{\beta\lambda\nu'}
-\Gamma^\mu_{\alpha\beta}D_{\mu\lambda\nu'}
-\Gamma^\mu_{\alpha\lambda}D_{\beta\mu\nu'}. 
\eqno (2.7)
$$

The Christoffel coefficients for our metric (1.1) read
$$
\Gamma^\alpha_{\beta\gamma} =
{1 \over 2}g^{\alpha\delta}\left( g_{\delta\beta,\gamma}+
g_{\delta\gamma,\beta}-
g_{\beta\gamma,\delta}\right)
= -{1 \over 2}{\epsilon \over a}
\left( \eta^{\alpha 0}\delta^0_\beta
\delta^3_\gamma + \eta^{\alpha 0}\delta^0_\gamma \delta^3_\beta-
\eta^{3\alpha}\delta^0_{\gamma}\delta^0_{\beta}\right).
\eqno (2.8)
$$
Since the connection coefficients, to first order in
$\epsilon$, are constant, we realize that the Ricci curvature tensor
vanishes to this order. On expanding (this is, in general, only an
asymptotic expansion)
$$
G_{\lambda\nu'} \sim G^{(0)}_{\lambda\nu'}+\epsilon \;
G^{(1)}_{\lambda\nu'} + {\rm O}(\epsilon^{2}), 
\eqno (2.9)
$$
we get
$$
D_{\beta\lambda\nu'}= \partial_\beta
G_{\lambda\nu'}-\Gamma^\mu_{\beta\lambda}G_{\mu\nu'}
=\partial_\beta
G_{\lambda\nu'}-\Gamma^\mu_{\beta\lambda}G^{(0)}_{\mu\nu'},
\eqno (2.10)
$$
so that
$$
Q_{\alpha\beta\lambda\nu'} =\partial_\alpha\partial_\beta
G_{\lambda\nu'}
-\partial_\alpha\left[\Gamma^\mu_{\beta\lambda}G^{(0)}_{\mu\nu'}\right]
-\Gamma^\mu_{\alpha\beta}\partial_\mu
G^{(0)}_{\lambda\nu'}
-\Gamma^\mu_{\alpha\lambda}\partial_\beta
G^{(0)}_{\mu\nu'},
\eqno (2.11)
$$
and eventually
$$ \eqalignno{
\cstok{ }_x G_{\lambda\nu'}&=
g^{\alpha\beta}\nabla_\alpha\nabla_\beta G_{\lambda\nu'}
=\left(\eta^{\alpha\beta}
+ \epsilon {z\over a} \delta^{\alpha}_0\delta^{\beta}_0 \right)
\nabla_\alpha\nabla_\beta\left[G^{(0)}_{\lambda\nu'}
+\epsilon
G^{(1)}_{\lambda\nu'} \right] \cr 
&=\eta^{\alpha\beta}\left[ \partial_\alpha\partial_\beta
G^{(0)}_{\lambda\nu'}+\epsilon
\partial_\alpha\partial_\beta
G^{(1)}_{\lambda\nu'}
-\Gamma^\mu_{\beta\lambda}G^{(0)}_{\mu\nu',\alpha}\right. \cr
& \left.-\Gamma^\mu_{\alpha\beta}
G^{(0)}_{\lambda\nu',\mu}-\Gamma^\mu_{\alpha\lambda}
G^{(0)}_{\mu\nu',\beta}\right]
-\epsilon {z\over a} \delta^{\alpha}_0\delta^{\beta}_0
\partial_\alpha\partial_\beta
G^{(0)}_{\lambda\nu'}.
&(2.12) \cr} 
$$
We therefore get, to first order in $\epsilon$,
$$
\cstok{ }^{0} G^{(0)}_{\mu\nu'}  = J^{(0)}_{\mu\nu'},
\eqno (2.13)
$$
$$
\cstok{ }^{0} G^{(1)}_{\mu\nu'} = J^{(1)}_{\mu\nu'},
\eqno (2.14)
$$
where
$$
J^{(0)}_{\mu\nu'}  \equiv -\eta_{\mu\nu}\delta(x,x'),
\eqno (2.15)
$$
$$
\epsilon J^{(1)}_{\mu\nu'}  \equiv {z\over a} \epsilon \left
({\eta_{\mu\nu} \over 2}+\delta^0_{\mu}\delta^0_{\nu}\right)\delta(x,x')
+2 \eta^{\rho\sigma}\Gamma^\tau_{\sigma\mu}
G^{(0)}_{\tau\nu',\rho} +
\eta^{\rho\sigma}\Gamma^\tau_{\rho\sigma} G^{(0)}_{\mu\nu',\tau}
-{z\over a} \epsilon G^{(0)}_{\mu\nu',00}, 
\eqno (2.16)
$$
with $\cstok{ }^0\equiv\eta^{\alpha\beta}
\partial_\alpha\partial_\beta=
-\partial_0^2+\partial_x^2+\partial_y^2+\partial_z^2$.

To fix the boundary conditions we note that, on denoting by
${\vec E}_{t}$ and ${\vec H}_{n}$ the tangential and normal components
of the electric and magnetic fields, respectively,
a sufficient condition to obtain
$$
\left . \vec{E}_{t} \right |_{S}=0,~~
\left . \vec{H}_{n} \right |_{S}=0,
\eqno (2.17)
$$
on the boundary $S$ of the device, is to impose Dirichlet
boundary conditions on 
$$
A_0(\vec{x}),A_1(\vec{x}),A_2(\vec{x})
$$
[14] at the boundary $z=0$, $z=a$. The boundary condition
on $A_3$ is determined by requiring that the gauge-fixing functional,
here chosen to be of the Lorenz type, should vanish on the boundary.
This implies
$$
\left . A^\mu_{;\mu} \right |_{S}=
0\Rightarrow \left . A^3_{;3} \right |_{S}
= \left . (g^{33}\partial_3
A_3-g^{\mu\nu}\Gamma^3_{\mu\nu}A_{3} )\right |_{S}=0 ~.
\eqno (2.18)
$$
To first order in $\epsilon$, these conditions imply the following
equations for Green functions:
$$
\left . G^{(0)}_{\mu\nu'} \right|_{S} = 0, ~~ \mu=0,1,2,
\; \forall \nu',
\eqno (2.19)
$$
$$
\left . \partial_3 G^{(0)}_{3\nu'} \right |_{S} = 0, ~~\forall \; \nu',
\eqno (2.20)
$$
$$
\left . G^{(1)}_{\mu\nu'} \right |_{S} = 0,
~~ \mu =0,1,2, \; \forall \nu',
\eqno (2.21)
$$
$$
\left . \partial_3 G^{(1)}_{3\nu'} \right |_{S}
= -{1 \over 2a} \left . G^{(0)}_{3\nu'} \right |_{S},
 ~~\forall \; \nu',
\eqno (2.22)
$$
hence we find that the third component of the potential $A_\mu$
satisfies homogeneous Neumann boundary conditions to zeroth order in
$\epsilon$ and inhomogeneous boundary conditions to first order.

Since $J^{(0)}_{\mu\nu'} $ is diagonal, by virtue of the
homogeneity of the boundary conditions, $G^{(0)}_{\lambda\nu'}$
turns out to be diagonal. On the contrary, $J^{(1)}_{\mu\nu'}$ has two
off-diagonal contributions: $J^{(1)}_{03} $ and $J^{(1)}_{30}$, so
that $G^{(1)}_{\mu\nu'}$ is non-diagonal. Let us write
down explicitly the expressions for the various components of
$J^{(1)}_{\lambda\nu'}$, i.e.
$$
aJ^{(1)}_{00'}  = {z \over 2}\delta(x,x') -z
G^{(0)}_{00',00}+ {1 \over 2}G^{(0)}_{00',3},
\eqno (2.23)
$$
$$
aJ^{(1)}_{03'}  = -G^{(0)}_{33',0},
\eqno (2.24)
$$
$$
aJ^{(1)}_{11'}  =   {z \over 2}\delta(x,x') -z
G^{(0)}_{11',00} - {1 \over 2}G^{(0)}_{11',3},
\eqno (2.25)
$$
$$
aJ^{(1)}_{22'}  =  {z \over 2}\delta(x,x') -z
G^{(0)}_{22',00} - {1 \over 2}G^{(0)}_{22',3},
\eqno (2.26)
$$
$$
aJ^{(1)}_{33'}  =  {z \over 2}\delta(x,x') -z G^{(0)}_{33',00}
- {1 \over 2}G^{(0)}_{33',3},
\eqno (2.27)
$$
$$
aJ^{(1)}_{30'}  = -G^{(0)}_{00',0}.
\eqno (2.28)
$$
Now we are in a position to  evaluate, at least formally (see
below), the solutions to zeroth and first order, and we get
$$
G^{(0)}_{\lambda\nu'} = \eta_{\lambda\nu'} \int{
{d\omega d^2k \over (2\pi)^3} e^{-i\omega(t-t')+ i
{\vec k}_{\perp}\cdot({\vec x}_{\perp}-{\vec x}_{\perp}')} }
g_{D,N}(z,z'),
\eqno (2.29)
$$
having defined
$$
g_{D}(z,z';\kappa) \equiv {\sin{\kappa(a
z_<)}\sin{\kappa(a-z_>)} \over \kappa\sin{\kappa
a} },~~~~~~~0<z,z'<a,
\eqno (2.30)
$$
$$
g_{N}(z,z';\kappa) \equiv - {\cos{\kappa(a
z_<)}\cos{\kappa(a-z_>)} \over \kappa\sin{\kappa a} },~~~~0<z,z'<a,
\eqno (2.31)
$$
where $D,N$ stand for homogeneous Dirichlet or Neumann boundary
conditions, respectively, $z_>~(z_<)$ are the larger (smaller)
between $z$ and $z'$, while ${\vec k}_{\perp}$ has components
$(k_x,k_y)$, ${\vec x}_{\perp}$ has components
$(x,y)$, $\kappa \equiv \sqrt{\omega^2-k^2}$, and
$$
G^{(1)}_{\mu \nu'}=\int{ {d\omega d^2k \over (2\pi)^3}
e^{-i\omega(t-t')+i{\vec k}_{\perp}\cdot({\vec x}_{\perp}
-{\vec x}_{\perp}')} \Phi_{\mu \nu'} },
\eqno (2.32)
$$
where the $\Phi$ components different from zero are written in Ref. [15].
A scalar field satisfies the same equations of the $22$ component of
the gauge field, hence we do not write it explicitly. In the following
we will write simply $G_{\mu\nu'}$ and $G$ for the Green
function of the gauge and ghost field, respectively.

We should stress at this stage that, in general, the integrals
defining the Green functions are divergent. They are well defined
until $x\neq x'$, hence we will perform all our calculations
maintaining the points separated and only in the very end shall we
take the coincidence limit as $x' \rightarrow x$ [16]. We
have decided to write the divergent terms explicitly so as to bear
them in mind and remove them only in the final calculations by
hand, instead of making the subtraction at an earlier stage.

Incidentally, we note that these Green functions satisfy the
Ward identity 
$$
G^\mu_{\; \nu';\mu}+G_{;\nu'}=0,~~~
G^{\mu \; \; ;\nu'}_{\; \nu'}+G^{;\mu}=0,
\eqno (2.33)
$$
to first order in
$\epsilon$ so that, to this order, gauge invariance is explicitly preserved
(the check being simple but non-trivial).
\vskip 0.3cm
\leftline {\bf 3. Energy-Momentum Tensor}
\vskip 0.3cm
\noindent
By virtue of the formulae of Sec. 2 we get, from the asymptotic expansion
$T_{\mu\nu'} \sim T^{(0)}_{\mu\nu'}+{\epsilon \over a} T^{(1)}_{\mu\nu'}
+{\rm O}(\epsilon^{2})$, 
$$
\langle T^{(0)\mu\nu'}\rangle = {1 \over 16\,a^4\,{\pi }^2}
\left( {\zeta}_{H}\left(4, {2\,a + z - {z'} \over 2\,a}\right) +
{\zeta}_{H}\left(4, {z'-z \over 2\,a}\right) \right)
{\rm diag}(-1,1,1,-3),
\eqno (3.1)
$$
where $\zeta_{H}$ is the Hurwitz $\zeta$-function
$\zeta_{H}(x,\beta) \equiv \sum_{n=0}^{\infty}(n+\beta)^{-x}$.
On taking the limit $z'\rightarrow z^+$ we get
$$
\lim_{z' \to z^+} \langle T^{(0)\mu\nu'} \rangle 
=\left({\pi^2 \over 720 a^4}
+\lim_{z' \to z^+} {1 \over \pi^2(z-z')^4}\right)
{\rm diag}(-1,1,1,-3),
\eqno (3.2)
$$
where the divergent term as $z' \rightarrow z$ can be removed by
subtracting the contribution of infinite space without bounding
surfaces [1], and in our analysis we therefore discard
it hereafter. The renormalization of the energy-momentum tensor in
curved spacetime is usually performed by subtracting the $\langle
T_{\mu \nu} \rangle$ constructed with an Hadamard or
Schwinger--DeWitt two-point function up to the fourth adiabatic
order [9,17]. In our problem, however, as we work
to first order in $\epsilon$, we are neglecting tidal forces and
therefore the geometry of spacetime in between the plates is flat.
Thus, we need only subtract the contribution to the energy
momentum tensor that is independent of $a$, which is the standard
subtraction in the context of the Casimir effect in flat spacetime.

In the same way we get, to first order in $\epsilon$:
$$ 
\lim_{z' \to z^+} \langle T^{(1)\mu\nu'}\rangle =
{\rm diag}(T^{(1)00},T^{(1)11},T^{(1)22},T^{(1)33}) 
+ \lim_{z' \to z^+} {\rm diag}\Bigr(-z'/\pi^{2}(z-z')^{4},0,0,0 \Bigr),
\eqno (3.3)  
$$
where
$$
T^{(1)00} = -{{\pi }^2 \over 1200\,a^3} +
{{\pi }^2\,z \over 3600\,a^4} +
{\pi \,\cot ({\pi \,z \over a})\,{\csc^{2} ({\pi
\,z \over a})} \over 240\,a^3},
\eqno (3.4)
$$
$$
T^{(1)11} = {{\pi }^2 \over 3600\,a^3} - {{\pi}^2\,z \over 1800\,a^4}
-{\pi \,\cot ({\pi \,z \over a})\,{\csc^{2} 
({\pi\,z \over a})} \over 120\,a^3},
\eqno (3.5)
$$
$$
T^{(1)22} = T^{(1)11},
\eqno (3.6)
$$
$$
T^{(1)33} = -{\left( {\pi }^2\,\left( a - 2\,z \right)
\right) \over 720\,a^4}.
\eqno (3.7)
$$
By virtue of the Ward identities for quantum 
electrodynamics, here checked up to first
order in $\epsilon$, the gauge-breaking part
of the energy-momentum tensor is found to be minus the ghost part, hence
we compute the second only. 
\vskip 0.3cm
\leftline {\bf 4. Casimir Energy and Force}
\vskip 0.3cm
\noindent
To compute the Casimir energy we must project the energy-momentum
tensor along the unit timelike vector $u$ with covariant
components $u_\mu=(\sqrt{-g_{00}},0,0,0)$ to obtain $\rho=\langle
T^{\mu\nu}\rangle u_\mu u_\nu$, so that 
$$ \eqalignno{ 
\rho &=
\left(1+\epsilon {z\over a}\right) \left[ -{\pi^2 \over 720
a^4}+{\epsilon \over a} \left(-{{\pi }^2 \over 1200\,a^3} +
{{\pi }^2\,z \over 3600\,a^4}  +
{\pi \,\cot ({\pi \,z \over a})\,{\csc^{2} ({\pi
\,z \over a})} \over 240\,a^3}\right) \right] \cr
&= -{\pi^2 \over 720a^4}+2 {g \over c^2}
\left(-{{\pi }^2 \over 1200\,a^3} -
{{\pi }^2\,z \over 900\,a^4}  +
{\pi \,\cot ({\pi \,z \over a})\,\csc^{2} 
({\pi\,z \over a}) \over 240\,a^3}\right)
+{\rm O}(g^{2}), 
&(4.1) \cr} 
$$
where in the second line we
have substituted $\epsilon$ by its expression in terms of $g$.
Thus, the energy stored in the Casimir device is found to be 
$$
E=\int{d^3\Sigma\sqrt{-g}\langle T^{\mu\nu}\rangle u_\mu
u_\nu}=-{\hbar c \pi^2 \over 720} {A \over a^{3}}
\left(1+{5 \over 2} {g a \over c^2} \right), 
\eqno (4.2)
$$
where $A$ is the area of the plates, $d^3\Sigma$ is the
three-volume element of an observer with four-velocity $u_\mu$,
Eq. (4.2) is expressed through a principal-value integral, and we
have reintroduced $\hbar$ and $c$.

In the same way, the pressure on the plates is given by
$$
P(z=0)={\pi^{2}\over 240} {\hbar c \over a^{4}}
\left(1+{2\over 3}{ga \over c^{2}}\right), \;
P(z=a)=-{\pi^{2}\over 240 } {\hbar c \over a^{4}}
\left(1-{2\over 3} {ga \over c^{2}}\right),
\eqno (4.3)
$$
so that a net force pointing upwards along the $z$-axis is obtained,
in full agreement with Eq. (8) in Ref. [6], with magnitude
$$
F={\pi^{2}\over 180}{A \hbar g\over c a^{3}}.
\eqno (4.4)
$$
The reader may wonder whether the pressure on the outer faces of the 
cavity may alter this result. A simple way to answer this question is to
imagine that our cavity is included into a surrounding cavity on both
sides. On denoting by $b$ the common separation between either
plates of the original cavity and the nearest plate of the surrounding 
cavity, and assuming that $b$ is such that 
$a/b <<1$, but still small enough so as to obtain
$gb/c^{2} << 1$, we see from Sec. 3 that the outer pressure on both plates
of the original cavity includes the same divergent contribution which acts
from within plus a finite contribution that becomes negligible for
$a/b <<1$. To sum up, the divergent contributions to the pressure 
from the inside and the outside of either plate cancel each other exactly, 
and one is left just with the finite contribution from the inside,
as given in Eq. (4.3).  

As a check of the result, it can be verified that the
energy-momentum tensor is covariantly conserved to first order in
$\epsilon$. To this order, the covariant conservation law implies
the following conditions: 
$$ 
\epsilon^0:  \langle
T^{(0)\mu\nu}\rangle _{,\nu}=0,
\eqno (4.5)
$$
$$
\epsilon: 
\langle T^{(1)ij} \rangle_{,j}=0~~(i=0,1,2), \; \;
{1 \over 2}\left(  \langle T^{(0)00}\rangle + \langle
T^{(0)33}\rangle \right)+ \langle T^{(1)33}\rangle_{,3} =0,
\eqno (4.6)
$$
that are indeed satisfied. Moreover, from the
previous expressions of the energy-momentum tensor the following
trace anomaly $\tau$ is obtained: 
$$ 
\tau= {\hbar g \over c a^{3}}\left(
{\pi^2 z \over 160 a}-{\pi \over 24}\cot{\left({\pi z
\over a}\right)}\csc^{2}{\left({\pi z \over a}\right)}\right).
\eqno (4.7)
$$ 
The volume integral of this density  exists as a
principal-value integral and is given by   
$$
\int \tau
~d^3\Sigma = {\pi^{2}\over 360}{\hbar g \over c a^{2}} \, A.
\eqno (4.8)
$$ 
The global, integrated  form (4.8) of the trace
anomaly  Eq. (4.7) is the new result with respect to the analysis
in Ref. [6]. It tends to zero at large separation $a$
between the plates.  This trace anomaly is therefore caused by
the presence of the boundaries, and then is of a different nature
from the usual trace anomaly which is encountered in curved
spacetimes without boundaries, which depends on the Riemann
curvature [11,16].
\vskip 0.3cm
\leftline {\bf 5. Concluding Remarks}
\vskip 0.3cm
\noindent
To the best of our knowledge, the analysis presented in this paper
represents the first study of the energy-momentum tensor for the
electromagnetic field in a Casimir cavity placed in a weak
gravitational field. The resulting calculations are considerably
harder than in the case of scalar fields. By using Green-function
techniques, we have evaluated the influence of the gravity
acceleration on the regularized energy-momentum tensor of the
quantized electromagnetic field between two plane-parallel ideal
metallic plates, at rest in the gravitational field of the earth,
and lying in a horizontal plane. In particular, we have obtained a
detailed derivation of the theoretical prediction according to
which a Casimir device in a weak gravitational field will
experience a tiny push in the upwards direction [6].
This result is consistent with the picture that the {\it negative}
Casimir energy in a gravitational field will behave like a {\it
negative mass}. Furthermore, we find a trace anomaly in Eq. (4.7)
proportional to the gravitational acceleration and vanishing for
infinite plates' separation, not previously worked out for a
Casimir device in a gravitational field. Our original results are
relevant both for quantum field theory in curved space-time, and
for the theoretical investigation of vacuum energy effects (see below).

We stress that in our computation we do not add by hand a mass
term for photons, unlike the work in Ref. [17], since
this regularization procedure breaks gauge invariance even prior
to adding a gauge-fixing term, and is therefore neither
fundamental nor desirable [12,16]. In agreement with
the findings of Deutsch and Candelas for conformally invariant
fields [18], we find that on approaching either wall, the
energy density of the electromagnetic field diverges as the third
inverse power of the distance from the wall. It is interesting to
point out that, in the intermediate stages of the computation,
quartic divergences appear in the contributions from the ghost and
the gauge breaking terms, which however cancel each other exactly.
The occurrence of these higher divergences in such terms is also
consistent with the results of Deutsch and Candelas, in view of
the obvious fact that ghost fields are not ruled by conformally
invariant operators. Unfortunately, a quantitative comparison with
their results is not possible because they {\it assume} a
traceless tensor, which is not the case in our problem where a
trace anomaly is found to arise.

Our results, jointly with the work in Refs. [6,19], 
are part of a research program aimed at studying the Casimir
effect in a weak gravitational field, with possible corrections
(albeit small) to the attractive force on the plates resulting
from spacetime curvature [20] (cf. the recent theoretical
analysis of quantum vacuum engineering propulsion in Ref.
[21]). Hopefully, these efforts
represent a first step towards an experimental verification of the
validity of the Equivalence Principle for virtual photons.
\vskip 0.3cm
\leftline {\bf Acknowledgements}
\vskip 0.3cm
\noindent 
The work of G. Bimonte and G. Esposito has been
partially supported by PRIN {\it SINTESI}. The work of L. Rosa has
been partially supported by PRIN {\it FISICA ASTROPARTICELLARE}.
\vskip 0.3cm
\leftline {\bf References}
\vskip 0.3cm
\noindent
\item{[1]}
M. Bordag, U. Mohideen, and V.M. Mostepanenko, 
Phys. Rep. {\bf 353}, 1 (2001).
\item{[2]}
G. Barton, J. Phys. A {\bf 34}, 4083 (2001).
\item{[3]}
N.G. van Kampen, B.R. Nijboer, and K. Schram, 
Phys. Lett. A {\bf 26}, 307 (1968).
\item{[4]}
N. Graham, R.L. Jaffe, V. Khemani, M. Quandt, M. Scandurra, 
and H. Weigel, Nucl. Phys. B {\bf 645}, 49 (2002).
\item{[5]}
M. Ishak, astro-ph/0504416; G. Mahajan, S. Sarkar, 
T. Padmanabhan, Phys. Lett. B {\bf 641}, 6 (2006).
\item{[6]}
E. Calloni, L. Di Fiore, G. Esposito, L. Milano, and L. Rosa,
Phys. Lett. A {\bf 297}, 328 (2002).
\item{[7]}
C. Misner, K.P. Thorne, and J.A. Wheeler, {\it Gravitation}
(Freeman, S. Francisco, 1973).
\item{[8]}
K.P. Marzlin, Phys. Rev. D {\bf 50}, 888 (1994).
\item{[9]}
S.M. Christensen, Phys. Rev. D {\bf 14}, 2490 (1976).
\item{[10]}
B.S. DeWitt, Phys. Rep. {\bf 19C}, 295 (1975).
\item{[11]}
B.S. DeWitt, `The Spacetime Approach to Quantum Field Theory', in
{\it Relativity, Groups and Topology II}, eds. B.S. DeWitt and
R. Stora (North--Holland, Amsterdam, 1984).
\item{[12]}
G. Bimonte, E. Calloni, L. Di Fiore, G. Esposito, L. Milano, and
L. Rosa, Class. Quant. Grav. {\bf 21}, 647 (2004).
\item{[13]}
L. Lorenz, Phil. Mag. {\bf 34}, 287 (1867).
\item{[14]}
G. Esposito, A.Yu. Kamenshchik, and G. Pollifrone, {\it Euclidean
Quantum Gravity on Manifolds with Boundary}, FTPHD,85,1, Fundamental
Theories of Physics, Vol. {\bf 85} (Kluwer, Dordrecht, 1997).
\item{[15]}
G. Bimonte, E. Calloni, G. Esposito, L. Rosa, 
Phys. Rev. D {\bf 74}, 085011 (2006).
\item{[16]}
R. Endo, Prog. Theor. Phys. {\bf 71}, 1366 (1984).
\item{[17]}
S.M. Christensen, Phys. Rev. D {\bf 17}, 946 (1978).
\item{[18]}
D. Deutsch and P. Candelas, Phys. Rev. D {\bf 20}, 3063 (1979).
\item{[19]}
R.R. Caldwell, {\it Gravitation of the Casimir effect and the 
cosmological nonconstant} (astro-ph/0209312).
\item{[20]}
F. Sorge, Class. Quant. Grav. {\bf 22}, 5109 (2005).
\item{[21]}
F. Pinto, J. Brit. Interpl. Soc. {\bf 59}, 247 (2006).

\bye